\documentclass{aa}

\usepackage{graphics,latexsym,psfig}
\def\degr{\hbox{$^\circ$\ }}

\def\arcsec{\hbox{$^{\prime\prime}$\ }}

\def\Ha{\hbox{H$\alpha$\ }}
\def\He2{\hbox{He{\sc ii} $\lambda$4686}}

\def\RL1{\hbox{{R$_{\rm L_{\rm 1}}$}}}
	
\begin{document} 
\title{SW Sextantis in an excited, low state} 
\author{P.J. Groot \inst{1,2}\thanks{CfA fellow}
\and R.G.M. Rutten\inst{3}
\and J. van Paradijs \inst{1,4}\,$^\dagger$}
\institute{Astronomical Institute `Anton Pannekoek'/ CHEAF, Kruislaan 403, 
1098 SJ, Amsterdam, The Netherlands
\and Harvard-Smithsonian Center for Astrophysics, 60 Garden Street,
Cambridge, MA 02138, USA
\and Isaac Newton Group of Telescopes, Apartado de Correos 321,
E-38700, Sta Cruz de La Palma, Islas Canarias, Spain
\and Physics Department, UAH, Huntsville, AL 35899, USA}
\date{Received date; accepted date} 
\offprints{pgroot@cfa.harvard.edu}

\abstract{ We present low-resolution spectrophotometric optical
observations of the eclipsing nova-like cataclysmic variable SW Sex,
the prototype of the SW Sex stars. We observed the system when it was
in an unusual low state. The spectrum is characterized by the presence
of strong He{\sc ii} and C{\sc iv} emission lines as well as the
normal single peaked Balmer emission lines. The radial temperature
profile of the disk follows the expected T$\propto$R$^{-3/4}$ only in
the outer parts and flattens off inside 0.5 times the white dwarf
Roche lobe radius.  The single peaked emission lines originate in a
region above the plane of the disk, at the position of the hot spot.\
}

\maketitle 
\section{Introduction \label{sec:intro}}
The SW Sex stars are a subclass of the eclipsing novalike (NL)
Cataclysmic Variables (CVs). They are classified on a number of
spectroscopic features, mainly the single peaked emission lines, the
large shifts between spectroscopic conjunction and photometric
mid-eclipse, the transient absorption evident in the emission lines
around photometric phase 0.5 and the shallow eclipse of the low
excitation lines of H{\sc i} and He {\sc i} (see Thorstensen et al., 1991;
Warner 1995). Their unusual spectroscopic behaviour has led to
intensive studies and a number of possible explanations for the SW Sex
`phenomenon': 
`wind' models (Honeycutt, Schlegel \&Kaitchuck, 1986), `overflow'
models (Hellier \&Robinson, 1994; Hellier 1996; Hellier 1998),
`magnetic' models (Williams 1989; magnetic accretion and Horne, 1999;
magnetic propellors) and `modified standard' models (Dhillon, Marsh \&
Jones, 1997, hereafter DMJ97). Although all of these are capable of
explaining a subset of the SW Sex phenonema, none show a conclusive
case of explaining all of the features listed above.

SW Sex itself is the prototype of the SW Sex stars and has been the
topic of many investigations. It was discovered in the Palomar-Green
Survey (Green, Schmidt \&
Liebert, 1986) as a UV-excess object with high excitation emission
lines (Green et al., 1982). Follow-up photometry showed it to be a
deeply eclipsing system with a 3.24 hr orbit (Penning et al.,
1984). We refer to DMJ97 for a recent, more detailed, description on the
observational history of SW Sex. 

From high-speed broad-band photometry Rutten, Van Paradijs \&
Tinbergen (1992, hereafter RvPT92) derived a radial dependence of the
temperature in the accretion disk that is not only flatter than seen
in other systems (see also RvPT92), but also flatter than predicted by
standard accretion disk theory (e.g. Frank, King \& Raine,
1992). This result, combined with the unusual behaviour of the
emission lines in SW Sex, prompted us to a study of the accretion disk
in SW Sex in more detail.  We obtained time series of
low-dispersion spectra of SW Sex with the 2.5m Isaac Newton Telescope
on the island of La Palma. 

\section{Data and Reduction \label{sec:reduction}}
Observations were made on the nights of 25 to 30 March, 1996 using the
Intermediate Dispersion Spectrograph (IDS) with the R300V
grating and a TEK 1k$\times$1k CCD, giving a dispersion of 3.3\AA\ 
per pixel over the 
wavelength range of 3600 \AA\ to 7000 \AA. In order to minimize
slit-losses and optimize the photometric quality of the data the slit
was opened to 2\arcsec. This set-up resulted in an effective
resolution of 8 \AA. A second star (30\arcsec\ SW of SW Sex) was also
placed on the slit to correct for slit-losses. 
We made time-series
with 90s integration and $\sim$60s dead-time per observation, giving
an effective time resolution of $\sim$150s or $\sim$ 1/80th
of an orbital period. Throughout the nights CuAr calibration spectra 
were taken for the wavelength calibration. A total of 282 spectra, 
covering 9 eclipses, were recorded. An overview of the data is given 
in Table\ \ref{tab:data}. 

\begin{table}
\caption[ ]{Overview of SW Sex observations March 1996 \label{tab:data}}
\begin{tabular}{llll}
Date 		& Start UT 	& End UT & No. Exposures\\[1mm]\hline\\[-1mm]
25/3/96 	& 22:17 	& 02:28  & 62\\
26/3/96 	& 20:20 	& 01:55  & 73\\
28/3/96         & 23:17		& 01:55  & 32\\
29/3/96		& 20:21		& 01:21	 & 70\\
30/3/96		& 22:16		& 02:20  & 45\\
\end{tabular}
\end{table}

The data reduction was carried out using the standard ESO-MIDAS reduction
package with additionally written software. All the observations were
debiased using the mean of the overscan region on each exposure. 
A run-averaged flatfield was constructed using internal Tungsten lamp
flats for the spectral profile and twilight skyflats for the spatial
profile. All spectra were optimally extracted (Horne, 1986) and
rebinned to a slightly oversampled wavelength-grid with 3\AA\ wide
bins. 
The wavelength calibration was accurate to 0.15 \AA. Time and
wavelength dependent slit losses were corrected for by applying the
variation in the comparison star brightness to SW Sex, while the
absolute slit losses were determined from comparing signal strengths
of the observations taken with the 2\arcsec wide slit with those taken
through a wide, 10\arcsec, slit. Flux
calibration was done by observing the spectrophotometric standard
Feige 34 (Massey et al., 1988) through the same set-up with a
10\arcsec wide slit. Based on Poisson statistics in the
extracted spectrum, each wavelength bin was assigned 
an error, which is propagated to the flux calibrated spectrum. 
The spectral slope of the comparison star did not correlate with
airmass and hence the slit losses due to differential refraction are
insignificant.


\section{Eclipse timing \label{sec:timing}}

Phase-folding of the nine observed eclipses, 
using the ephemeris of DMJ97, gave a phase-offset
of minimum light in the photometric light curves of $\sim$0.006 in
phase.  We redetermined the ephemeris of SW Sex using the eclipse
timings as given in DMJ97 and by making Gaussian fits to our own
eclipse light curves (see Table\ \ref{tab:times} for all eclipse
timings). For the existing data we used an error of 1$\times$10$^{-4}$
days on the time of minimum light. For our own data we used an error
of 0.5$\times$10$^{-4}$ days, as estimated from the Gaussian fits.
 A linear regression yields the revised
orbital ephemeris given in Eq.\ \ref{eq:eph}.

\begin{table}
\caption[ ]{Times of minimum light for SW Sex \label{tab:times}}
\begin{tabular}{llr}
HJD (mid-eclipse) & Cycle Number & O-C \\
(+2\,440\,000) & ($N$) & (s) \\[1mm]\hline\\[-1mm]
4339.65087 &     0 &  --26.33 	\\    
4340.73055 &     8 &  --33.44    \\ 
4348.82649 &    68 &  --23.50    \\
4631.92758 &  2166 &  --42.92     \\
4676.86195 &  2499 &  --6.74   \\
4721.79636 &  2832 &  --12.73    \\
7566.56813 & 23914 &  8.20     \\
7615.41619 & 24276 &  13.33   \\
7615.55065 & 24277 &  28.26    \\
7616.49516 & 24284 &  48.41    \\
7618.51922 & 24299 &  61.44    \\
7619.46374 & 24306 &  39.39    \\
7620.40856 & 24313 &  17.35   \\
7621.48834 & 24321 &  10.24      \\
7622.43257 & 24328 &  30.38    \\
7921.32167 & 26543 &  --13.94     \\
7921.45633 & 26544 &  0.98     \\
7950.19842 & 26757 &  17.28   \\
7950.33321 & 26758 &  32.22    \\
8306.30100 & 29396 &  17.89   \\
8306.43599 & 29397 &  32.82    \\
10168.452377& 43196 & --58.15      \\
10168.586651& 43197 & --43.22      \\
10169.396799& 43203 & --38.01      \\
10169.531017& 43204 & --23.07      \\
10171.554764& 43219 & 32.14    \\
10172.364589& 43225 & 37.35    \\
10172.500202& 43226 & --32.09      \\
10173.443720& 43233 & 72.42     \\
10173.579514& 43234 & 2.98  
\end{tabular}
\end{table}

\begin{table}
\caption[]{SW Sex system parameters as used for the eclipse mapping
\label{tab:sys} }
\begin{tabular}{ll}
Period  & 11658.67 s\\
M$_{\rm WD}$ & 0.44 M$_{\odot}$\\
M$_{\rm sec}$ & 0.3 M$_{\odot}$\\
Inclination & 79\degr\\
Distance     & 290 pc\\
\end{tabular}
\end{table}	

\begin{eqnarray}
HJD_{\rm mid\_eclipse}& =& 2\,444\,339.650574(36) \nonumber \\
		      &  & + 0.1349384411(10) \times N
\label{eq:eph}
\end{eqnarray}

\begin{figure*}[htb]
\centerline{\psfig{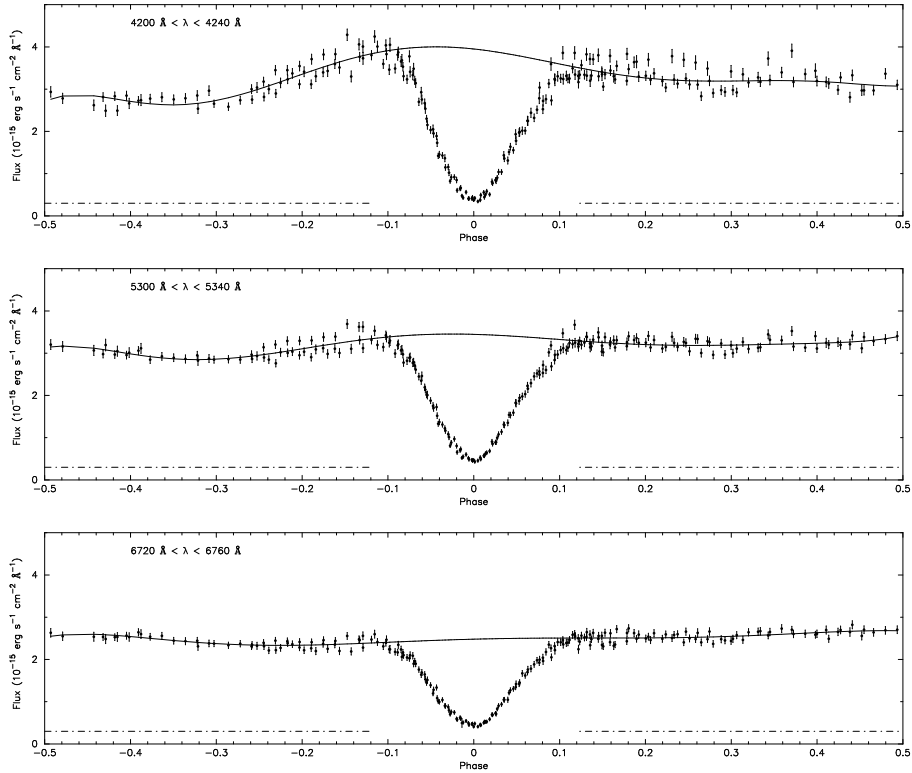}}
\caption[]{The spectrophotometric light curve of SW Sex between {\mbox 4200\AA\
$< \lambda <$ 4240 \AA\ (top)}, 
\mbox{5300\AA\ $< \lambda <$ 5340\AA\ (center)}, and 
\mbox{6720\AA\ $< \lambda <$ 6760 \AA\ (bottom)}. A pronounced orbital hump is
seen in the blue light curve, which gradually diminishes and is gone at
the reddest wavelengths. The 7th order polynomial fits to the light
curves outside eclipse are shown as the full lines. \label{fig:light}}
\end{figure*}

\section{Continuum light-curves \label{sec:contlight}}

For the spectrophotometry we used seven of nine observed
eclipses. The remaining two (of the night of March 28 and the second
eclipse of the night of March 30) were not used because the S/N levels
of the fainter, secondary star on the slit were too low to make a
reliable slit correction. 
This leaves a total of 220 spectra of
SW Sex to be used. 

From the spectrophotometry
light curves can be extracted at any of the observed wavelength regions. 
Figure\ \ref{fig:light} shows the continuum light curves between 
$\lambda\lambda$ 4200--4240,
5300--5340 and 6720--6760 \AA\ as examples. 
Apart from the eclipse, the light curves show the presence of an 
orbital hump that disappears towards longer wavelenghts.

Comparison of the continuum light curves with previously published
data shows a similar amount of eclipse profile asymmetry as in 
DMJ97, Penning et al. (1984) and Ashoka et al. (1994) for the $\lambda$
4200--4240 \AA\ light curve. 

The elementary eclipse-mapping program, as will be applied in 
Sec.\ \ref{sec:mapping}, assumes that all variation in the light curve occurs
during eclipse. We therefore need to apply a correction to take away
any variations outside elipse. We have applied
a polynomial fit to the phases smaller than --0.12 and larger than
0.12. Trial fits with different order polynomials showed that a 7th
order polynomial returned the smallest residuals, independent of
wavelength. 

\section{Spectral Eclipse Mapping \label{sec:mapping}}

Eclipse mapping uses the information contained in the shape of the
eclipse light curve to reconstruct the light distribution on the
accretion disk. For a general description of the method we refer to
Horne (1985). We have used the same program as was used by Rutten et
al. (1994) for the analysis of UX UMa, which is a maximum-entropy
based optimization routine, as described in Horne (1985; see also Gull
\& Skilling, 1983) For the reconstruction of the accretion disk in SW
Sex we used the basic, thin, flat disk approach, as was also used in
Rutten et al. (1992), who also show that the temperature profile is
not affected by the assumptions about the shape or thickness of the
disk. No correction for interstellar extinction was deemed necessary
since IUE spectra show no presence of the 2200 \AA\ bump (RvPT92).

\subsection{Light curve selection and preparation}

We have split the total wavelength range covered by our spectra
(3650--7000 \AA) in 79 wavelength bins, each 40 \AA\ wide, with the
exception of the spectral lines, which were taken as one bin each
(with variable width).

The errors which were assigned to each spectral bin on the basis of
Poisson statistics were propagated to the light curves.
SW Sex was very
stable during the six nights of our observations and no matching of
the individual light curves was required. 
A small amount of random variation is seen in the 
blue part of the spectrum, but this diminishes towards the red, 
and is therefore most likely dominated by intrinsic
flickering. These variations will cause the $\chi^2$-based
eclipse mapping routine to reconstruct spurious fine structure in the
resulting map. To avoid this we have increased the formal errors by up
to 40\% for the bluest wavelengths. In the red part of the spectrum
this adjustment was not more than 10\%.

For the reconstructions as presented here, we used a 51$\times$51
pixel map, with the phase interval --0.20 $< \varphi < $0.20 as
input. 
We have set no {\it a priori} limit to the size of the
accretion disk.The light curve zero-point was taken as a free
parameter in the fit to account for any uneclipsed light (RvPT92).

\subsection{Reconstructed intensity distribution}

\begin{figure}[htb]
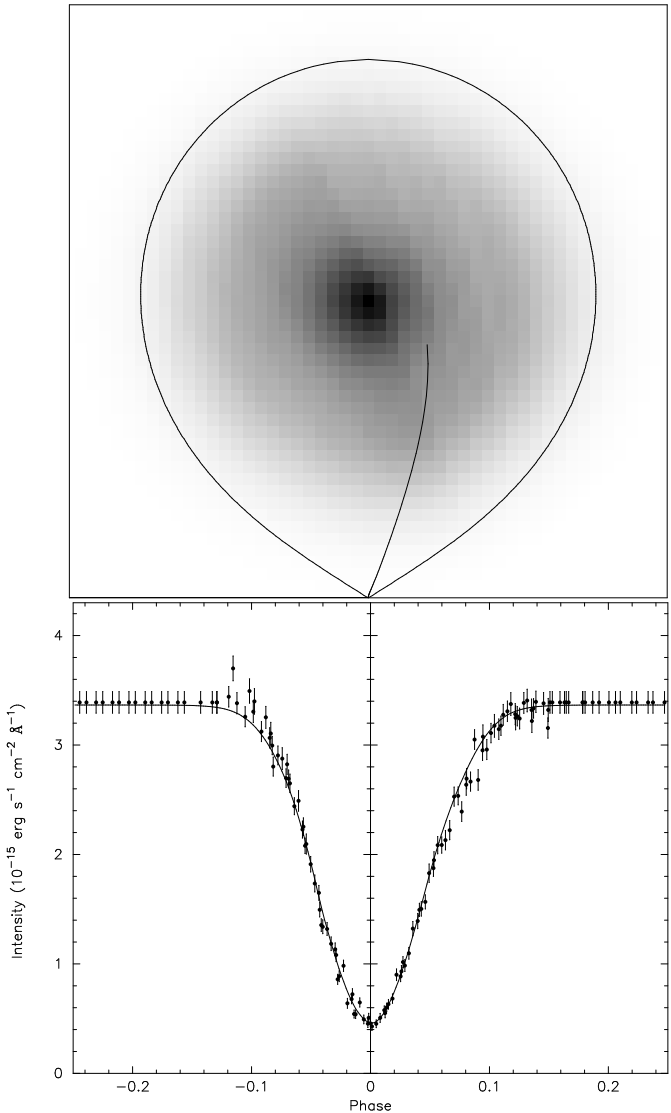

\hfill{\psfig{figure=mapyellow.ps,width=8cm,angle=-90}}
\centerline{\psfig{figure=lightyellow.ps,width=8.8cm,angle=-90}}
\caption[]{The reconstructed accretion disk for the wavelength region
between 5490 \AA\ and 5530 \AA\ (top). The bottom panel shows the 
input (dots) and reconstructed (line) lightcurves. \label{fig:mapyellow}}
\end{figure}

In Figs.\ \ref{fig:mapyellow} we
show the resulting eclipse map in the middle part of our spectrum. 
The map shows a
clear dominance of the white dwarf and surroundings, 
which is reconstructed as the bright region in the middle. On the
scale used here the size of the WD corresponds to $\sim \frac{1}{4}$ pixel. 
We see that the emission is dominated by the WD and surroundings.
In the blue maps a hot-spot contribution is seen, whose strength
diminishes towards the red, consistent with the light curve behaviour.

\subsection{Size of the accretion disk}

To measure the size of the accretion disk we use the same measure as
in RvPT92, namely the point where the intensity has fallen to 10\% of
the central intensity.  At 4060 \AA\ we measure R$_{\rm d} \simeq$ 0.7
\RL1, where \RL1 is the distance between the inner Lagrangian point
(L$_1$) and the white dwarf. At azimuth angles that are pointing
towards the hot-spot region (as seen from the WD) the apparent size of
the disk reaches up to 0.9 \RL1.  At 5510 \AA\ the disk reaches up to
0.65-0.7 \RL1, slightly depending on azimuth angle, and at 6350 \AA\
the 10\% level is reached at 0.75 \RL1, independent of azimuth
angle. The azimuth dependence with wavelength reflects the diminishing
influence of the hot-spot in SW Sex. The disk radii measured here are
slightly larger than measured by RvPT92. 

\subsection{Accretion disk spectrum}

\begin{figure}[htb]
\centerline{\psfig{figure=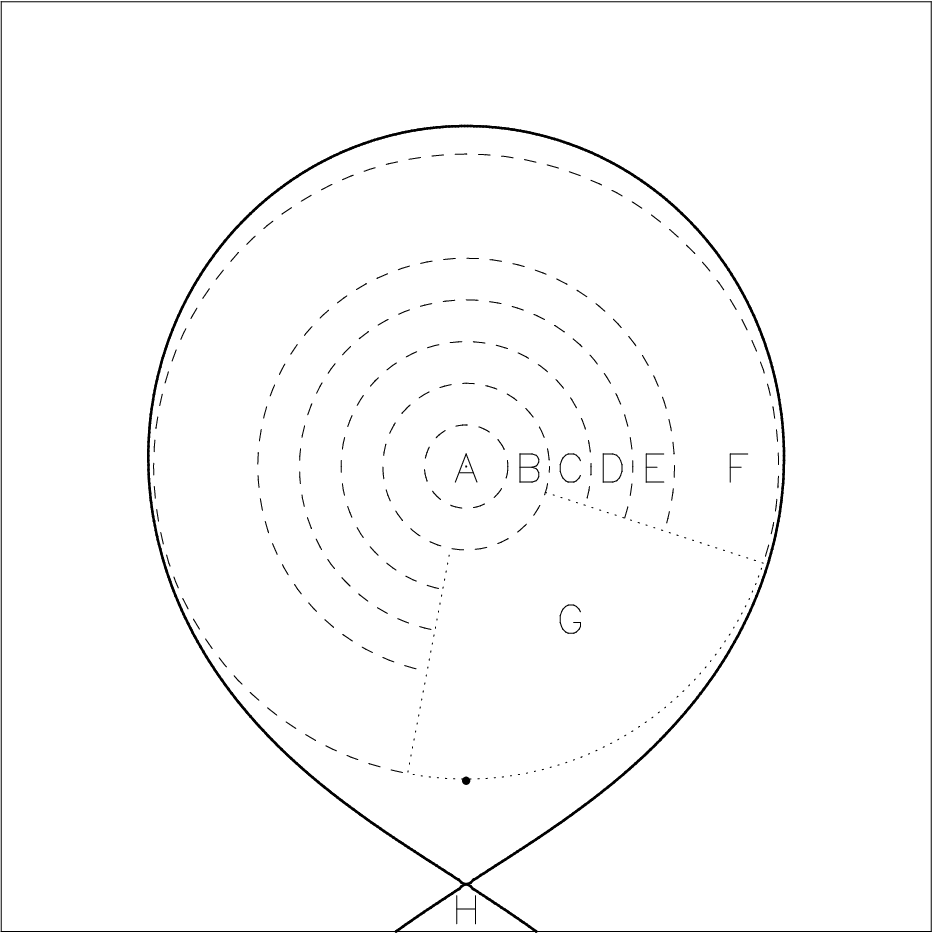,angle=-90,width=8cm}}
\caption[]{To reconstruct the accretion disk spectrum the WD
Roche-lobe has been divided in 7 segments, labeled `A' through `G' and
located, respectively, between 0-0.1, 0.1-0.2, 0.2-0.3, 0.3-0.4, 0.4-0.5,
0.5-0.75\RL1. The segments further out than 0.2\RL1 are split up 
between the part between phase 0.03 and 0.8 (section `F') and the 
hot-spot part between phase 0.8 and 0.03 (section `G'). 
The component of uneclipsed light is represented by segment `H', 
here tentatively located at the secondary. \label{fig:rings}}
\end{figure}

\begin{figure*}[htb]
\centerline{\psfig{figure=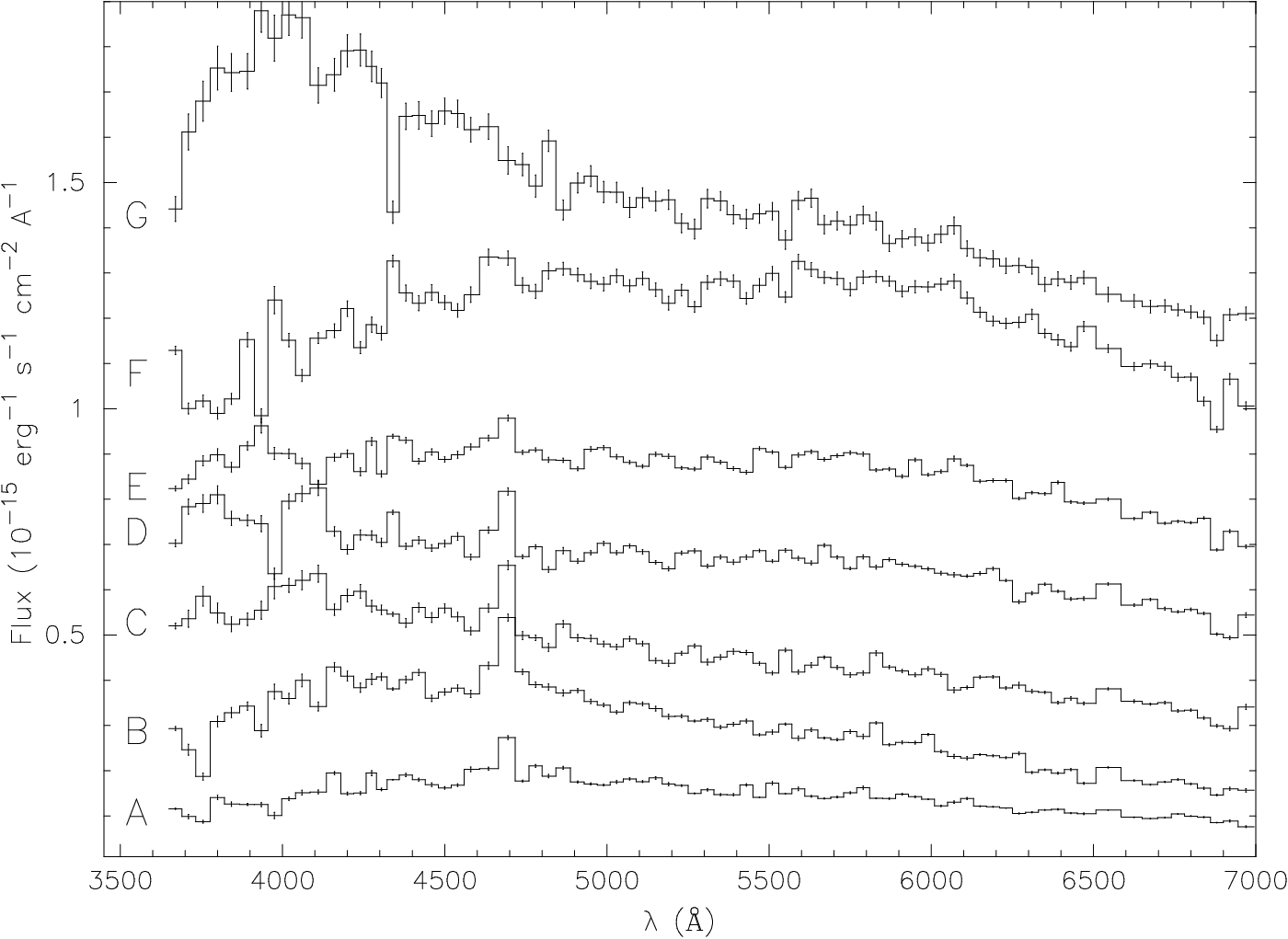,angle=-90,width=12cm}}
\caption[]{Reconstructed accretion disk spectrum in the regions A
through G as indicated in Fig.\ \ref{fig:rings}. To separate the
spectra from each other shifts have been applied of respectively
--0.1, 0.05, 0.25, 0.46, 0.55, 0.85 for the regions A through
G. Fluxes have been calculated as the summed flux in each
segment. \label{fig:radialspectrum}}
\end{figure*}

The reconstructed intensity maps allow us to deduce the spectrum at any given
point in the disk. Since the eclipse-mapping procedure smooths the
disk in the azimuthal direction and since we are mainly concerned with
the radial profile of the disk, we have used a set of concentric rings
(shown in Fig.\ \ref{fig:rings}) and a hot-spot region. The
reconstructed spectra are shown in Fig.\ \ref{fig:radialspectrum}. 
There is spectral variation in the radial
direction, going from a relatively blue spectrum in the middle of the
disk to a redder spectrum near the edges of the disk. The hot-spot
region (`G') is clearly much bluer than the other section of the outer
ring (`F'). Comparing the spectral evolution seen here with the one
that was found in UX Uma by Rutten et al. (1994), we see that the
changes in SW Sex are much more moderate and less pronounced than in UX
Uma, whose temperature profile does follow the
T$\propto$R$^{\frac{-3}{4}}$ relation as predicted by the theory of
steady state accretion disks (see e.g. Frank, King \& Raine; 1992)

\subsection{Uneclipsed light \label{sec:offset}}

\begin{figure}[htb]
\centerline{\psfig{figure=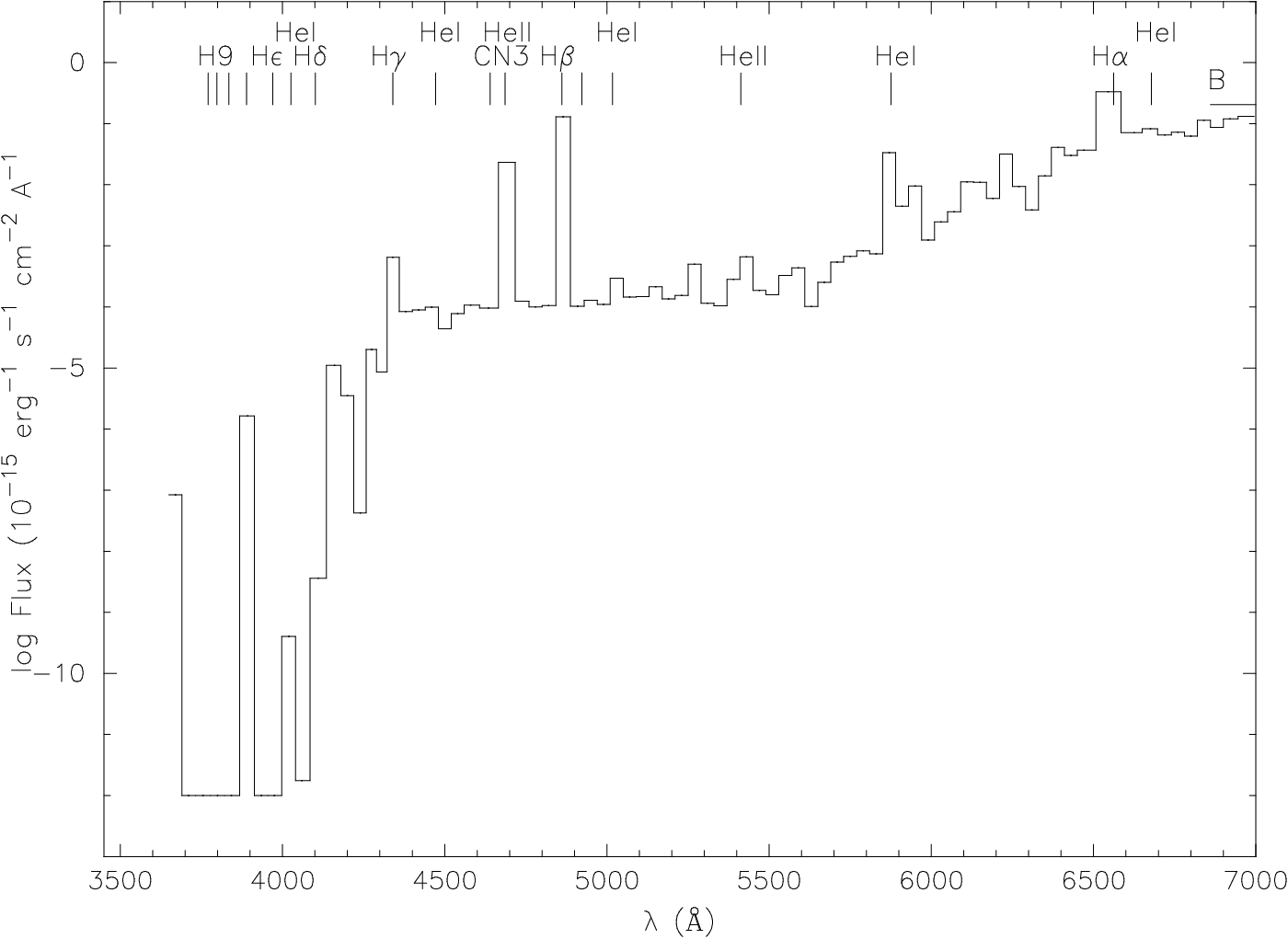,angle=-90,width=8.8cm}}
\caption[]{The reconstructed spectrum of the uneclipsed light
component.The continuum levels $<$4200 \AA\ are computational noise.
 \label{fig:offset}}
\end{figure}

Any uneclipsed component to the light curve could produce spurious
results in the final eclipse map (see RvPT92). For that reason 
the uneclipsed light component was incorporated as a free parameter
in the optimization routine. 
The spectrum of this
uneclipsed component is shown in Fig.\ \ref{fig:offset}. 
It prominently shows the Balmer series in emission. A well known
feature of SW Sex systems is the partial eclipse of the Balmer lines
and the occurence of these lines in emission in the uneclipsed
component is consistent with this. The uneclipsed component is most
likely to emanate from either the secondary star or from outside the
orbital plane. The secondary star in SW Sex has not been detected. 
The slope of the uneclipsed spectrum as reconstructed here is
indicative of a late M-type secondary, which suggests (some of) the
light could represent the secondary.  

\subsection{Accretion disk temperature distribution \label{sec:disktemp}}

We reconstruct a temperature distribution
on the accretion disk by making Planck curve fits to the spectrum at
each pixel on the disk, assuming the complete disk is optically thick and
radiating as a blackbody. The flux received on
Earth is:

\begin{equation}
f = \frac{{N}{A} \cos i \sum_{j=1}^{N} \sigma
{T}_j^4}{4 \pi d^2},
\label{eq:flux}
\end{equation}
with $N$ the number of pixels on the disk, $A\cos i$, the projected size
of each pixel, $T_j$ the effective temperature of each pixel and $d$ the
distance to the source. 

\begin{figure}[htb]
\centerline{\psfig{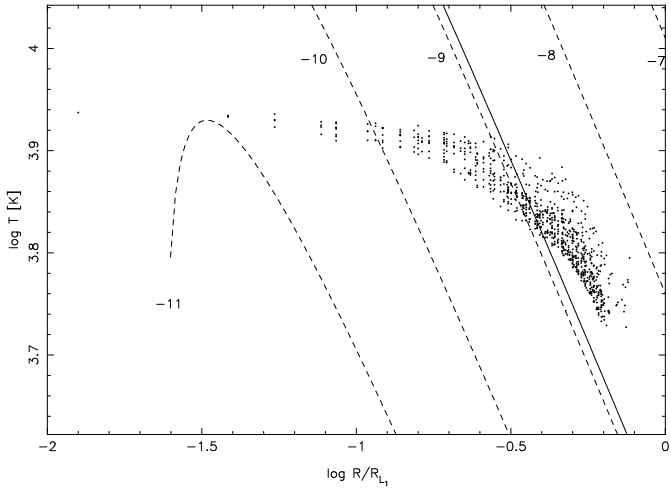}}
\caption[]{The derived $\dot{M}$ profile as function of the radial
distance from the white dwarf. The solid line shows the averaged
fitted mass-transfer rate. The dashed lines show the expected profile
on the basis of standard accretion disk theory in units of 10$\log$ of
the mass transfer rate in M$_{\odot}$
yr$^{-1}$  (i.e. `--10' is 10$^{-10}$ M$_{\odot}$
yr$^{-1}$). \label{fig:mdot}}
\end{figure}

The distance to SW Sex is not well known. An uncertain value of 250 pc
is quoted by Patterson (1984), derived from the H$\beta$ equivalent
width and the continuum shape. RvPT92 derived 450 pc from blackbody
fits to the inner accretion disk in their four colour photometry and
the absence of any feature of the secondary.  Including the distance
as a free parameter in our blackbody fits, we derive a distance to SW
Sex of 290 pc. Using the value of 450 pc as found by RvPT92 gave
clearly incorrect blackbody fits at all points in the disk. 
Using 290 pc for the distance,
we obtain a temperature map of the accretion disk of SW Sex, of
which the radial cut is shown in Fig.\ \ref{fig:mdot}. Note that any
uncertainty in the distance affects the temperatures but less so the
radial dependence. 

\begin{figure*}[htb]
\centerline{\psfig{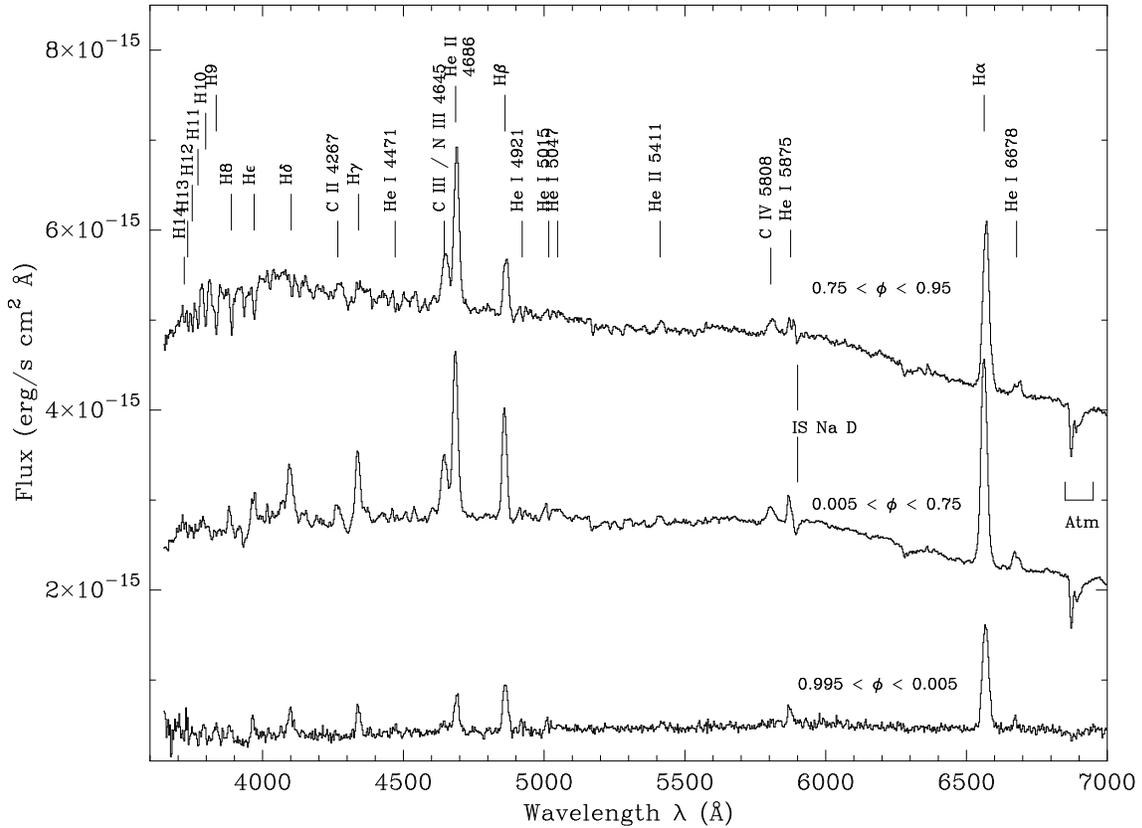}}
\caption[ ]{The average spectrum of SW Sex at phase intervals 0.995
$<\varphi<0.005$ (bottom), 0.005 $<\varphi<$ 0.75 (middle) and
0.75$<\varphi<$0.95 (top, offset by +2.0$\times$10$^{-15}$ erg s$^{-1}$ 
cm$^{-2}$ \AA$^{-1}$). Major lines are indicated. 
\label{fig:ave}}
\end{figure*}

The current results confirm those of RvPT92 that the radial temperature
dependence does not follow the standard, steady state prediction, but
deviate at radii $<0.5\RL1$. Although Smak (1994) argued that the flat
temperature profile is due to obscuration of the inner parts of the
disk, this would require a disk flare angle of more than 11\degr,
which appears rather large (see also Rutten 1998). 

\section{Spectral line behaviour}

In the spectral eclipse mapping we have used the spectral lines as
single wavelength bins. We will now investigate in further detail
the spectral line behaviour. 

\begin{figure*}[htb]
\centerline{\psfig{figure=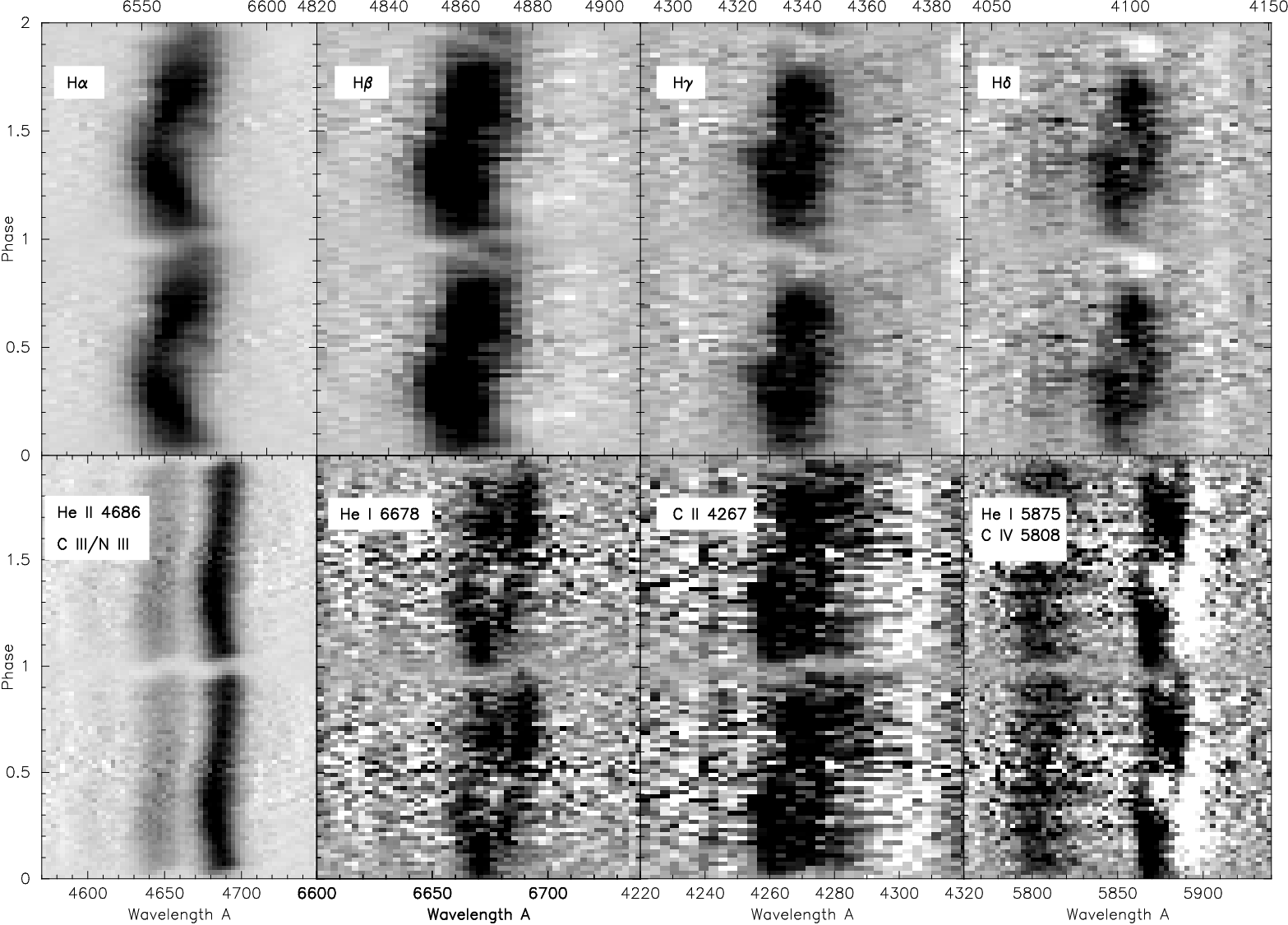,width=16cm,angle=-90}}
\caption[]{Trailed spectra of H$\alpha$, H$\beta$, H$\gamma$ and
H$\delta$ (top, left to right), and \He2/ C{\sc iii}, N{\sc iii} 
$\lambda$4645, He{\sc i} $\lambda$6678, C{\sc ii}
$\lambda$4267 and He{\sc i} $\lambda$5875 and C{\sc iv} $\lambda$5808
(bottom, left to right). 
The He{\sc i} $\lambda$5875 line is badly affected by the Na{\sc i}
doublet next to it.\label{fig:trail}}
\end{figure*}

\subsection{Average spectrum\label{sec:ave}}

Before we `zoom in' on the spectral lines we first show the average
spectrum of SW Sex during our run in Figure\ \ref{fig:ave}. 

The out-of-eclipse (0.005$<\varphi<$0.75) shape of the spectrum of SW
Sex (Fig.\ \ref{fig:ave}, middle curve) is typical for SW Sex stars:
strong single peaked Balmer, He{\sc ii} and N{\sc iii}/C{\sc iii}
emission on top of a flat or slightly blue continuum. However, there are a
number of marked differences with previous spectroscopic studies of SW
Sex (Green et al., 1982; Penning et al., 1984; Williams, 1989 and
DMJ97):
\begin{itemize}
\item The ratio of \He2 over H$\beta$ is larger than unity.
\item The He{\sc i} lines are unusually weak.
\item The presence of C{\sc iv} $\lambda$5805 and He{\sc ii} $\lambda$5411. 
\end{itemize}
These three unusual features show that the emission lines are formed
in a site with a higher than usual ionization level. 

Furthermore, the system is approximately 1.2 mag fainter than when
observed by Penning et al. (1984), Honeycutt et al. (1986) 
and RvPT92. The depth of the eclipses (see Sect.\
\ref{sec:contlight}) is not significantly different (1.9-2.0 mag) from
previous epochs.

\subsection{Multiple emission components \label{sec:multemis}}
We have binned the data in 50 phase bins and subtracted the underlying
continuum by making a linear fit to adjacent wavelengths. The trailed
spectra of the main emission lines are shown in Fig.\ \ref{fig:trail}.

It is clear that all lines show a complex behaviour, with
multiple emission and absorption components present. The low excitation
Balmer lines (\Ha, H$\beta$) and \He2 are dominated by an S-wave, with
variable emission strength. However, the higher excitation Balmer
lines, the He{\sc i} lines and the C{\sc ii} $\lambda$4267 line have multiple
components. Although the low-resolution of our data defeats an unambigious
interpretation, the behaviour of the He{\sc i} $\lambda$6678 and
$\lambda$5875 lines suggests two separate S-waves
that intertwine. Our preference for this, over the explanation of one
S-wave with variable absorption components superposed on it, is given
by the difference in the emission profile width at phase 0.4 compared
to phase 0.1. 
The He{\sc i} $\lambda$5875 is badly affected by the interstellar Na D doublet
at $\lambda$ 5890 \AA.  

\subsection{Radial-velocity curves of the main S-wave component
\label{sec:radvel}}

\begin{figure}[htb]
\centerline{\psfig{figure=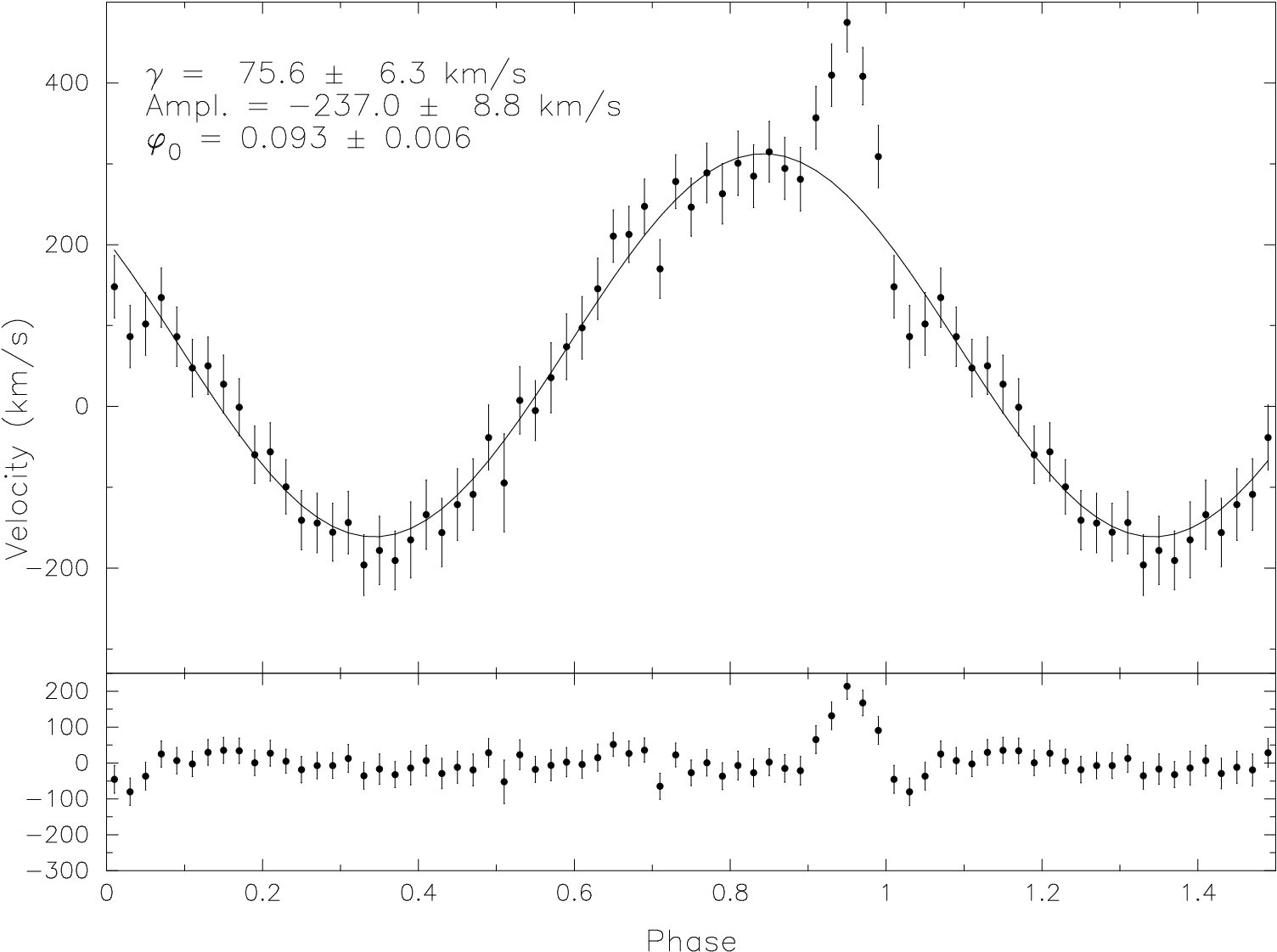,width=8.8cm,angle=-90}}
\centerline{\psfig{figure=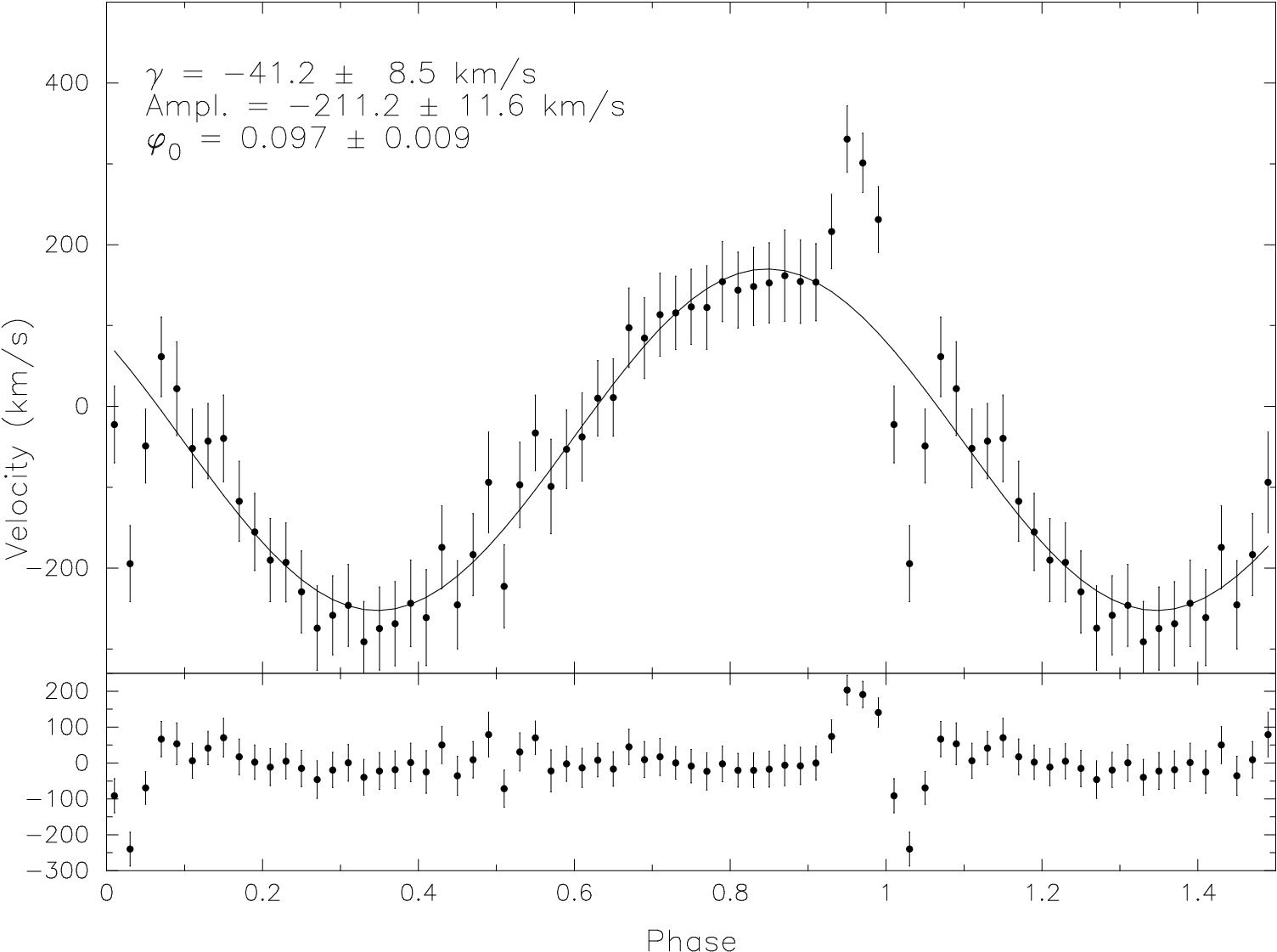,width=8.8cm,angle=-90}}
\caption[]{Radial velocity curve of \Ha (top) and \He2 (bottom). Full
lines show the best sinusoidal fit to the data. Residuals are given in
the lower panels. 
\label{fig:radvelHaHeII}}
\end{figure}

The radial-velocity curves of the main component visible in H$\alpha$
and \He2 are shown in Fig.\ \ref{fig:radvelHaHeII}.  These were
determined by fitting a single Gaussian to the line profiles at phase
intervals of 0.02. The Gaussian fit is effectively determined by the
flanks of the emission lines.  In both lines the main component has
its red-to-blue crossing at $\varphi=0.09$. This indicates that the
emission is not symmetrically centered on the WD, since in that case
the red-to-blue crossing would coincide with superior conjunction of
the white dwarf, i.e. at $\varphi$=0. This phase lag of the emission
lines is a well known characteristic of SW Sex stars (see
e.g. Thorstensen et al. 1991). The phase lags are in good agreement
with the results from DMJ97.

The 0.09 phase offset indicates that the emission site is not
on the line of centers, between the center-of-mass (CoM) and the
white-dwarf, but at a slight angle to it. (See Fig.\ \ref{fig:rocheover}
for a graphic summary). The identical phase lag and amplitude of the
radial velocity curve indicate that both \Ha as well as \He2 originate in
the same location. 

If we take the average of the measured values for the two spectral
lines as the (resultant) velocity vector 
of the emitting material and correct for the orbital
inclination ($i$=79\degr), we get $v_{\rm res}$ = 230 km/s for the
emitting material. 

Both radial velocity curves show a rotational disturbance around
mid-eclipse. This rotational disturbance is caused by the progressive
obscuration of first the blue-shifted gas, and later the
red-shifted gas in the rotating accretion disk. 
The occurence of a rotational disturbance shows that part of the
emission originates in the accretion disk. 

For both lines maximum redshift of the rotational disturbance occurs
at $\varphi \sim$0.95, which coincides with the moment of ingress of
the white dwarf and the hot-spot region. Maximum blue-shift, however,
is reached for both lines at phase, $\varphi \sim$0.03. This is too
early to be explained by an emission site in the plane of the disk if
Roche geometry holds. This indicates that, at $\varphi\sim$0.03, the main
emission site becomes visible again and dominates the emission profile
over disk component. 

\begin{figure*}[htb]
\centerline{\psfig{figure=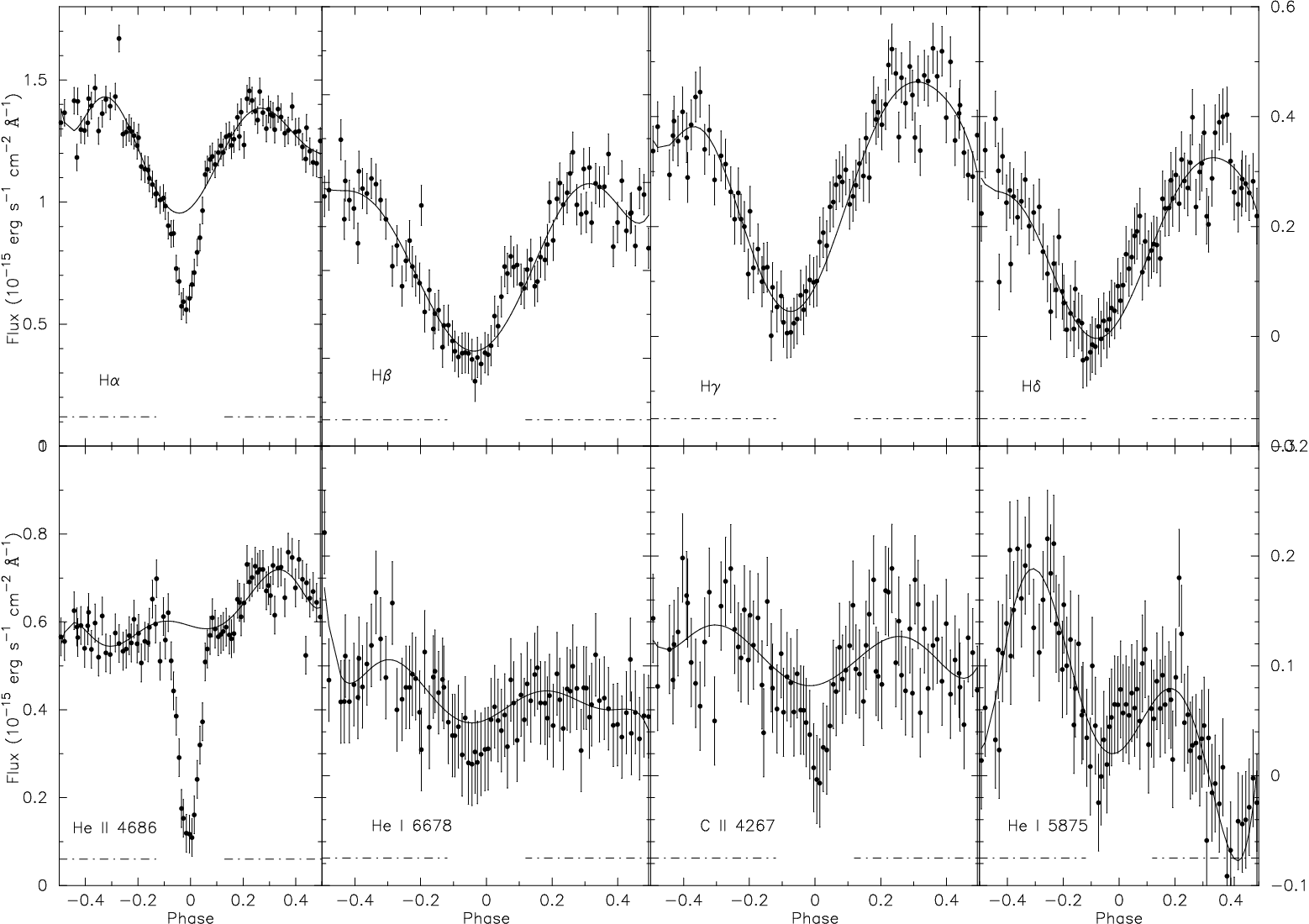,angle=-90,width=16cm}}
\caption[]{The spectrophotometric light curves of the lines presented
in Fig.\ \ref{fig:trail}. Data have been rebinned to 100 phasebins to
allow for meaningfull continuum subtraction. The solid line shows the
7th order polynomial fit to the phase intervals outside -0.12 and 0.12
(indicated by the dashed-dotted line). The scale for H$\beta$ is 0 to
1 $\times$10$^{-15}$ erg s$^{-1}$ cm$^{-2}$ \AA$^{-1}$, for H$\gamma$
and H$\delta$ it is given on the top right hand side of the plot, for
He I $\lambda$6678, C{\sc ii} $\lambda$4267 and He{\sc i} $\lambda$
5875 it is given on the bottom right side of the figure.
\label{fig:linelight}}
\end{figure*}

The observed (phase-lag, radial velocity amplitude)-pair is the
resultant of a vector addition of two components: the orbital
velocity, which is directed perpendicular to the line joining the
emission site and the center-of-mass, and the flow velocity of the
material in the emission site around the white dwarf.  If we know the
position of the emission site, we can calculate the orbital velocity
of material at that particular distance of the WD and since we
also know the resultant velocity vector from the radial velocity
curve, we can determine the true gas-flow velocity of the material
around the WD (see Sect.\ \ref{sec:emissionvel}).

We have not attempted to obtain radial velocity curves for any of
the other lines presented in Fig.\ \ref{fig:trail} because of the
multicomponent nature of these lines, as presented and discussed in
Sect.\ \ref{sec:multemis}.

\subsection{Line light-curves \label{sec:linelight}}

Line light-curves for the lines presented in Sec.\ \ref{sec:multemis}
were extracted by summing all data in the relevant wavelength regions
and subtracting from these a linear fit to the adjacent continuum
(Fig.\ \ref{fig:linelight}). As was already evident in the trailed
spectra (Fig.\ \ref{fig:trail}), the line light-curves show
substantial variations outside eclipse. The distorted profiles are
very similar to those that have been seen before in SW Sex (DMJ97).

The light curve of \He2 is very similar to
that of the continuum, and is dominated by the eclipse. In the other light
curves presented in Fig.\ \ref{fig:linelight} only \Ha and C{\sc ii}
$\lambda$4267 show the presence of an eclipse. The Balmer series light
curves are dominated by a broad V-shaped feature that is centered on
$\varphi \sim$0.95. 

\section{The Hot Spot \label{sec:hotspot}}

\subsection{Position of the hot-spot \label{sec:hotpos}}

\begin{figure}[htb]
\centerline{\psfig{figure=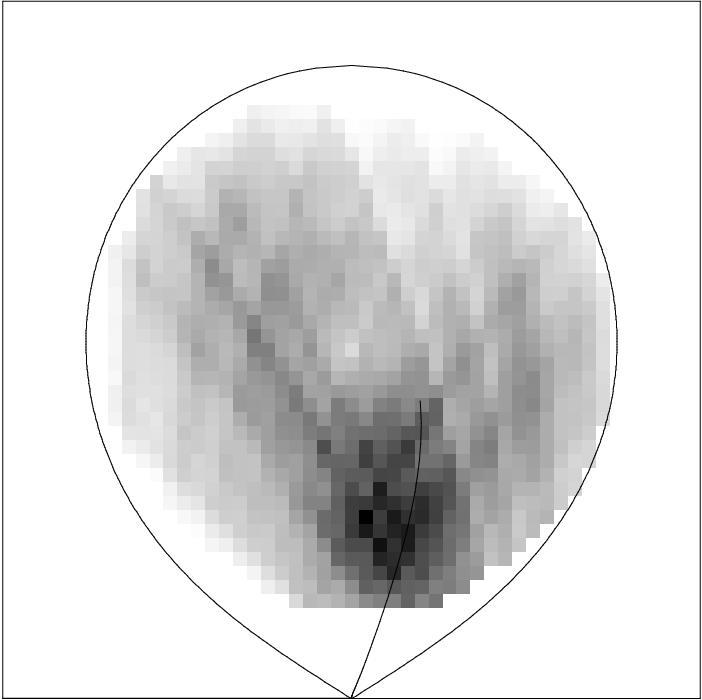,width=8cm,angle=-90}}
\caption[]{The reconstructed hot-spot map between 3800\AA\ and
4140\AA. The map was made by summing the maps in this wavelength
region and dividing it by the map at 6350\AA\ to obtain solely the
hot-spot part. The map shows that the hot-spot is located at 0.4-0.6
\RL1, and has azimuthal extent of 0.93$<\varphi<$1.0 in phase. 
\label{fig:hotmap}}
\end{figure}

To determine the position of the hot-spot we have added up the eclipse
maps between 3800\AA\ and 4140\AA, where the hot-spot continuum
spectrum is the brightest. This sum map is then divided by a map at
red wavelengths to obtain the ratio of the two. This ratio map is
shown in Fig.\ \ref{fig:hotmap} and brings out the location of the
hot-spot at $r_{\rm hs}$ = 0.4-0.6 \RL1 and azimuth angles 
0.93$<\varphi<$1.0.  The
azimuthal extent is somewhat uncertain because of the smearing in the
azimuthal direction as applied in the eclipse mapping method.

\subsection{Hot-spot absorption line spectrum \label{sec:hotspec}}

\begin{figure*}[htb]
\centerline{\psfig{figure=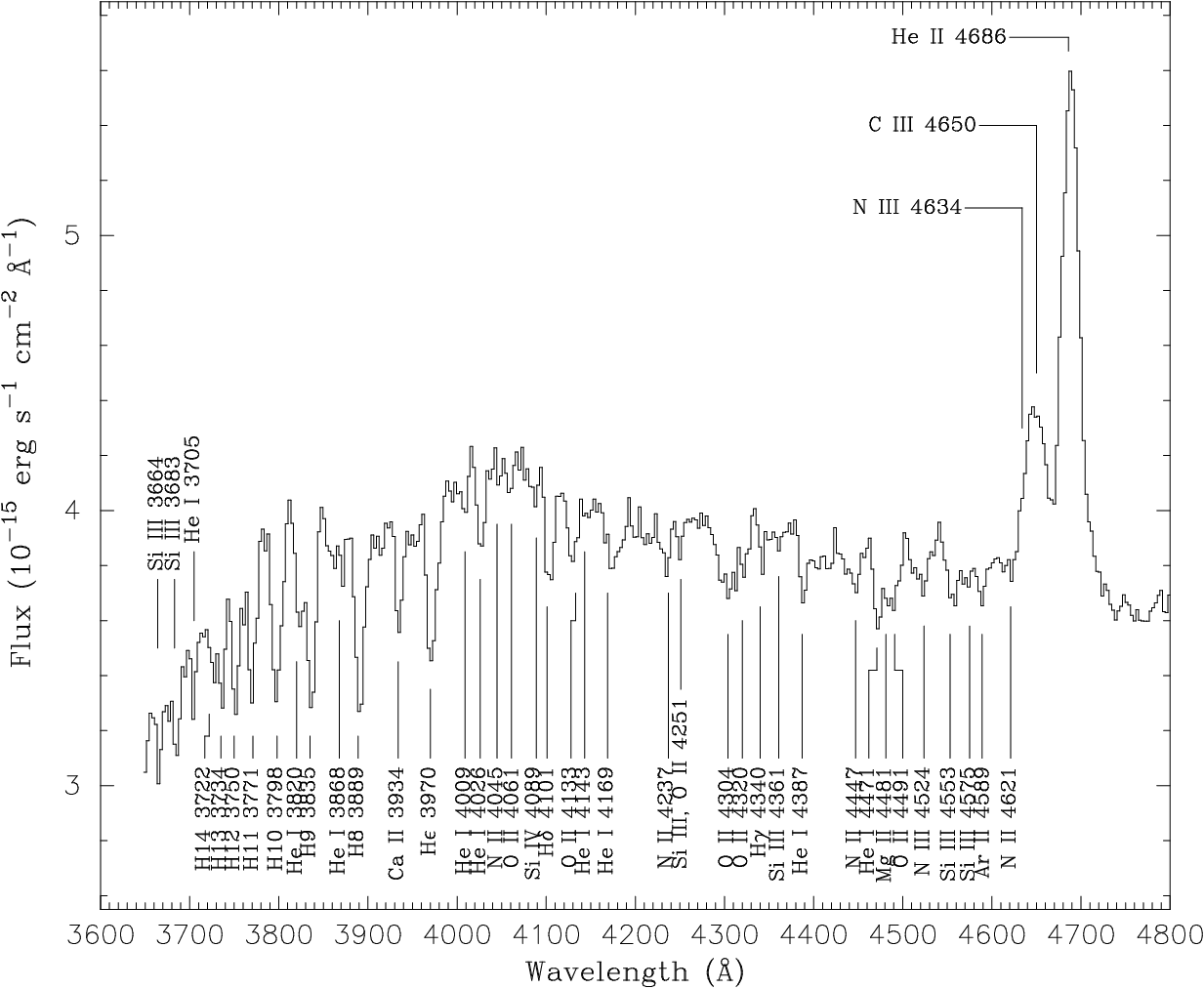,angle=-90,width=16cm}}
\caption[]{The blue part of the average spectrum at
0.75$<\varphi<$0.95. The absorption line identifications show the
spectrum to be identical to a B0-type spectrum, indicating absorbing
material temperatures of 19\,000 K $<$ T$_{\rm abs}<$ 25\,000
K. Line identifications have been taken from Van Kerkwijk et
al. (1995). 
\label{fig:hotspec}}
\end{figure*}

During the phases that the hot-spot is visible, 0.75$<\varphi<$0.95,
the blue spectral shape changes dramatically with respect to the
spectrum during the rest of the orbit. The higher Balmer lines change
from general emission between phases 0.005$<\varphi<$0.75 to
absorption, visible up to H14, as shown in Fig.\ \ref{fig:hotspec}.
 At the same time, the higher excitation
lines of \He2, C{\sc ii} $\lambda$4267 and the 
C{\sc iii}/N{\sc iii} $\lambda$4645 complex do {\it not} change in
strength compared to the continuum and each other. 

The ensemble of absorption lines as present in the hot-spot spectrum
can be identified as those occuring in an early B-type spectrum (e.g.
the B0 type star HD77581/Vela X-1; Van Kerkwijk et al., 1995). Typical
temperatures in stellar photospheres of B0-type stars range between
19\,000 K $<$ T$_{\rm abs}<$ 25\,000 K, depending on local
gravity. Since we expect a relatively low local gravity in the
hot-spot region ($g\sim 0.1 g_{\odot}$, comparable to a giant's
photospheric gravity; Marsh 1988), the temperature will be on the
lower end of this range. Deducing a more exact determination of the
temperature of this absorption region is compounded by two factors:
the low-resolution of the data and the fact that most of the
absorption lines that are normally used for spectral classification of
OB-type stars (He{\sc i}, He{\sc ii}, C{\sc ii}, C{\sc iii}) are in
emission in SW Sex. However, we can set a tentative lower limit
of $\sim$15\,000 K (corresponding to a B2 spectral type), from the
fact that the Si{\sc iv} $\lambda$4089 \AA\ line is visible and the
Mg{\sc ii} $\lambda$4481 \AA\ line is not visible.

\subsection{Absorption line spectrum light curve}

\begin{figure}[htb]
\centerline{\psfig{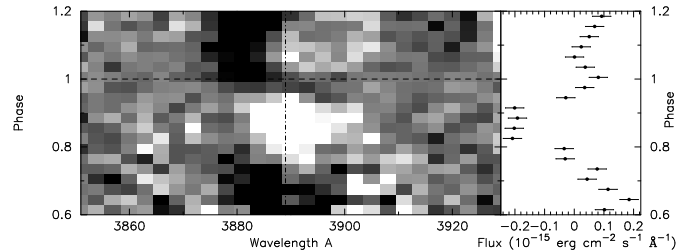}}
\caption[]{Trailed spectrogram of the H8 ($\lambda$3889) line (left) and
its lightcurve (right) from phase 0.6 to 1.2. The horizontal dashed
line shows mid-eclipse according to Eq.\ \ref{eq:eph}, the vertical
dashed line is the central wavelength. 
\label{fig:trailH8}}
\end{figure}

For this we have constructed the trailed
spectrum and light curve of H8 (Fig.\ \ref{fig:trailH8}). We have
chosen this line because it is clearly present in the absorption line
spectrum of Fig.\ \ref{fig:hotspec} and is relatively well isolated,
which allows a reliable continuum subtraction.  We see in Fig.\
\ref{fig:trailH8} that the light curve of H8 does {\it not} resemble
that of the lower Balmer series. Instead of a wide V-shape, it is
almost U-shaped and the absorption only extends from
0.82$<\varphi<$0.94.

\subsection{The hot spot continuum spectrum \label{sec:hotcont}}

Although we have deduced in the previous paragraph that the hot-spot
region must contain gas at temperatures exceeding 15\,000 K, this
appears to be at odds with the eclipse mapping results given in Sect.\
\ref{sec:disktemp}, where we see that the temperature in the hot-spot
region does not exceed 10\,000 K. Here we will investigate whether the
temperature derived from the absorption spectrum, T$_{\rm abs}$ is the
same as that derived from the hot-spot continuum radiation, T$_{\rm
spot}$.

To this end we take the spectrum as reconstructed in the spectral
eclipse mapping in region 'G' in Fig.\ \ref{fig:rings}. 
Since this region will also contain parts of the
accretion disk which are not affected by the accretion stream impact,
and since we expect the hot-spot region to be the hottest part of this
region, we will concentrate on the blue part of the spectrum
($\lambda<$5000 \AA) to derive the hot-spot continuum temperature.

We have used the spectral catalogue of Jacoby, Hunter \& Christian
(1984) to compare the hot-spot region 'G' spectrum with that of a wide
range of spectral types and luminosity classes. As mentioned above we expect
the effective gravities in the hot-spot region to be
similar to a giant's atmosphere, and we have therefore restricted
ourselves to spectra of luminosty class III, although the differences
in luminosity class are almost indistinguishable at the low resolution
of the reconstructed hot-spot spectrum. 

\begin{figure}[h]
\centerline{\psfig{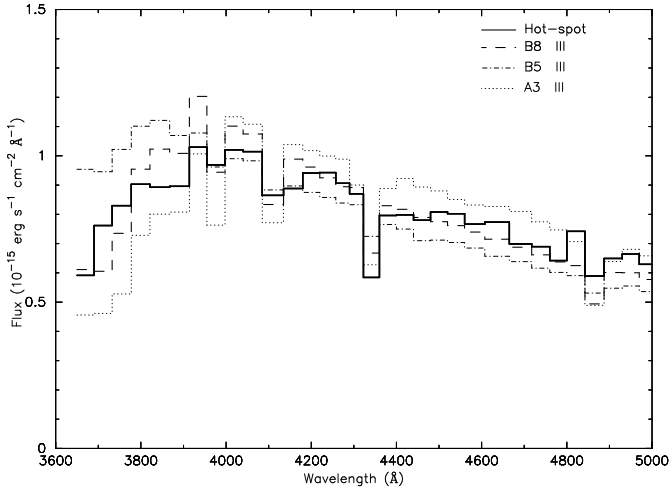}}
\caption[]{The reconstructed hot-spot spectrum, 
compared with three standard star spectra from Jacoby, Hunter
\&  Christian (1984). 
\label{fig:hotref}}
\end{figure}

In Fig.\ \ref{fig:hotref} we compare our hot-spot
spectrum (solid line) with the B8\,III spectrum of HD28696, rebinned to
the same spectral resolution and scaled to the same flux (thick dashed
line). 
Also shown are a B5III and an A3III
spectrum. The B5III spectrum underestimates the Balmer jump
considerably, and hence the hot-spot 
continuum must be cooler than B5III (T$_{\rm spot}<$12\,000 K). 

Consequently, the gas, which we see in the absorption line spectrum as
presented in Sect.\ \ref{sec:hotspec} is of a higher temperature 
(with a lower limit of T$_{\rm abs}>$15\,000 K) 
than the gas that emits the hot-spot continuum radiation, 
for which we can put a very conservative upper limit of 
T$_{\rm spot}<$12\,000 K. 
If the Balmer jump in our hot spot continuum spectrum were to be
underestimated (e.g. due to emission in the higher Balmer lines) the
discrepancy between the temperature deduced from the absorption lines
and the continuum would become even worse.  

\section{Discussion}

\subsection{Emission region velocities \label{sec:emissionvel}}

\begin{figure}[htb]
\centerline{\psfig{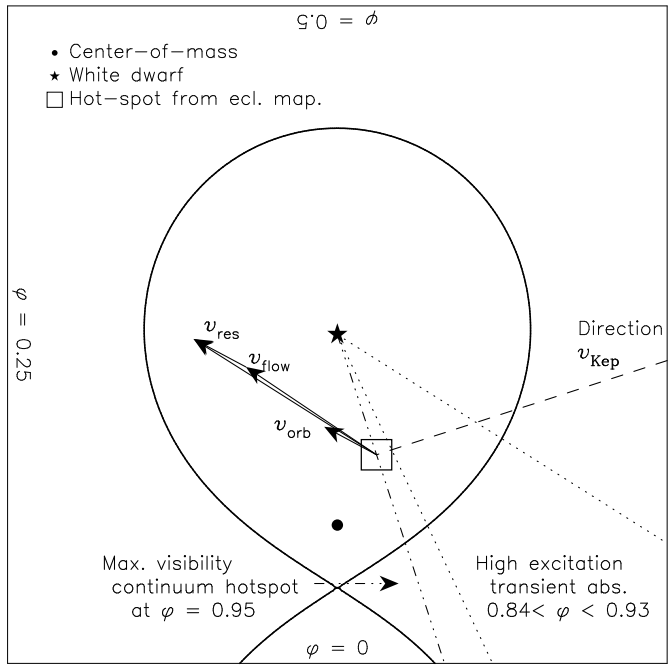}}
\caption[]{A schematic view of the Roche lobe of the white dwarf in SW
Sex. The orbital velocity, $v_{\rm
orb}$, at the hot spot and the observed velocity vector, $v_{\rm res}$,
are almost parallel. The gas flow velocity vector $v_{\rm flow}$,
deduced from this is $\sim$170 km/s.
\label{fig:rocheover}}
\end{figure}

With the localization of the hot spot from the eclipse mapping and the
\Ha and \He2 radial velocity curves we can reconstruct the velocity
vector of the material in the line emission region. In Fig.\
\ref{fig:rocheover} we show a schematic picture of the white dwarf
Roche lobe with the velocity vectors included. The orbital velocity
($v_{\rm orb}$) at the position of the hot spot, using the system
parameters of SW Sex as listed in Table\ \ref{tab:sys}, is 60 km/s. We
see that the resultant velocity vector as measured from the radial
velocity curves is directed in almost the same direction as the
orbital velocity, but has a much higher value. Combined we deduce that
the flow velocity of the material in the emission line region is of
the order of 170 km/s and is directed almost in the same direction as
the orbital velocity. If the gas in the emission region is connected to
the gas flow in the accretion disk, we would have expected a more or
less Keplerian flow with a velocity of the order of $\sim$500 km/s. We
can see in Fig.\ \ref{fig:rocheover} that both the direction and the
magnitude of the Keplerian velocity vector are very different from
those observed. We therefore conclude that the gas at the emission
site is decoupled from the gas in the accretion disk.

Note also that a disk overflow component, directed in the direction
of the continuation of the accretion stream (roughly directed towards
$\varphi\sim$0.6), will not be enough to explain the resultant $v_{\rm
flow}$ component.

\subsection{The hot spot structure} 

The radial velocity curves as well as the light curves of the higher
Balmer lines point towards the fact that these lines are formed in the
vicinity of the hot spot.  The correct hot-spot structure will have to
combine four results into one picture: the continuum temperature
T$_{\rm spot}$, the absorption line temperature T$_{\rm abs}$, the
V-shaped line light curves and the U-shaped line light curves of the
absorption line spectrum. 

We can reconcile these by adopting a hot-spot structure that consists
of two components: a photospheric component with T$_{\rm spot}$ and a
chromospheric component with T$_{\rm abs}$. The photospheric component
is responsible for the continuum radiation as derived from spectral
eclipse mapping, but is also responsible for the V-shaped light curves
of the lower Balmer series. The strength of the Balmer absorption
lines in a stellar spectrum reaches a maximum in late B, early A-type
stars. The hot-spot spectrum will therefore contain deep photospheric
Balmer absorption lines. The strength of these lines in the observed
spectrum will depend on our viewing angle towards the hot-spot: coming
into view at $\varphi\sim$0.7, reaching a maximum around
$\varphi\sim$0.95 and disappearing from view again at
$\varphi\sim$0.2. Since at the same time there is a more extended
region that produces Balmer line emission, and which is to first order
independent on phase, we see the effect of the hot spot absorption
lines as a decrease in emission in the line light curve.

It can also be seen in Fig.\ \ref{fig:linelight} that the amount of
emission actually increases again during primary eclipse (see the
excess emission with respect to the fit in H$\beta$, H$\gamma$ and
H$\delta$). This is due to the photospheric site (the
hot spot) being eclipsed more than the emission-line site. We conclude
that the emission lines are formed in an extended halo around the
hot-spot region. 

The observed absorption-line spectrum is only visible during the phase
interval 0.82$<\varphi<$0.94 and displays the more U-shaped light
curve (Fig.\ \ref{fig:trailH8}). 
This component is caused by veiling by material above the hot
spot (the chromospheric component with temperature T$_{\rm abs}$) of
continuum radiation from the hot innermost parts of the accretion
disk. Only during the phase interval stated above is this hot spot
chromosphere `back-lit' by radiation that is emitted by
regions that are hot enough to cause the gas to show up in
absorption. The sharp begin of the absorption shows that this hot
region is very small itself (most likely only the white dwarf and very
direct surroundings). At $\varphi$=0.94 the eclipse by the secondary star
terminates our line of sight towards this hot-spot chromosphere.
  
\subsection{The structure of SW Sex}

\begin{figure*}[htb]
\centerline{\psfig{figure=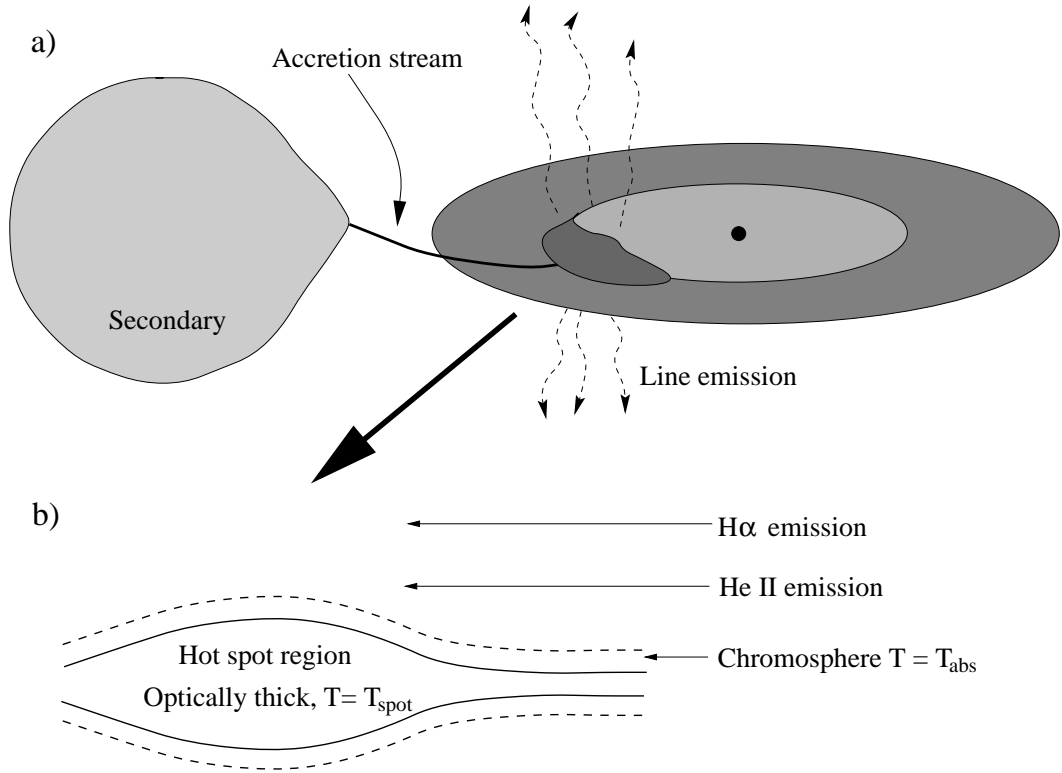,width=14cm,angle=-90}}
\caption[]{Schematic view of SW Sex (top) and its hot-spot region
(bottom).
\label{fig:swsex}}
\end{figure*}

The results from the spectral eclipse mapping and hot spot
structure derived above combines into the following picture of SW
Sex (see also Fig.\ \ref{fig:swsex}): the accretion disk is in a quasi
steady state with mass transfer rate, $\sim$10$^{-9}$ M$_{\odot}$/yr,
in the outer parts, but the temperature profile flattens inside a
radius of $\sim$0.5 \RL1. The outer
part of the disk is dominated by the hot spot region, which is also
the center of a larger, vertically extended, region which is the
emission site of the main emission lines, as evidenced by the phase
lags and radial velocity curves of the main S-wave components in \Ha
and \He2. 

What is causing both the temperature disturbance in the inner disk and
the extended hot-spot halo is unclear. We view it as
outside the scope of the current paper to go into details on the
possibilities. We would, however, like to point out a resemblance of
the characteristics of SW Sex as discussed here with the old nova V
Per, and we will shortly discuss the possibility of SW Sex stars being
intermediate polars.

\subsection{V Per: a twin to SW Sex?}
A very interesting case and a possible twin of SW Sex is the little
studied old nova V Per. Shafter \&  Abbott (1989) have shown that this
is an eclipsing CV in the period gap (P$_{\rm orb}$=2.57 hr), that
shows single peaked emission lines, and most interesting, \He2 (with a
strength $>$H$\beta$), and He {\sc ii} $\lambda$5411 in emission; two
characteristics that are also seen in our data of SW Sex. Since, to
our knowledge, no phase-resolved spectra of V Per have been published, 
nothing is known of the relative phasing of its emission lines with
respect to the continuum eclipses. However, Wood, Abbott \&  Shafter
(1992) have shown from broad-band eclipse light curves, that the disk
temperature profile is very similar to that found here for SW
Sex. Wood et al. explain this behaviour by the absence of the inner
part of the accretion disk and speculate that this could be the result
of a magnetic field, which would make V Per a possible intermediate
polar. A phase-resolved spectroscopic study of this system is required
to understand the nature of this system and its relation to the
subclass of SW Sex stars. If it is found to show the `classic' SW Sex
phenomena (phase lags, transient absorption), it would extend the
orbital range of SW Sex stars considerably downwards, and if V795 Her
(Casares et al., 1996) is indeed a SW Sex star, be the second SW Sex
star in the period gap.

\subsection{SW Sex stars as intermediate polars ?}

It has been suggested that SW Sex stars (and also V Per) are intermediate
polars at high inclination.  It is clear that we need a source
of high energy photons in SW Sex to explain the strength of the \He2, He{\sc
ii} $\lambda$5411 and C{\sc iv} $\lambda$5808 lines, and the weakness of the
He{\sc i} lines in our spectrum of SW Sex. Although this source
of high energy photons could be a magnetic white dwarf, these line
strengths are also
the {\it only} suggestions for a magnetic white dwarf. No periodic
oscillations, as seen in almost all polars and intermediate polars, are
found in the optical in any of the SW Sex stars. Hard X-ray
emission is not detected from SW Sex stars, no cyclotron features
have been observed and in V1315 Aql no polarization is detected
(Dhillon \&  Rutten, 1995). Although none of these are
clearly arguing against a magnetic white dwarf (since exceptions in
magnetic systems can be found for all of them), the lack of any
of the characteristics leaves the question of the source of X-ray flux
open. Above all, we have seen that the
emission site of the \He2 radiation is {\it not} connected to the
white dwarf, but to the hot-spot region. We conclude that
although there is not much support for a magnetic white dwarf as the
source of high-energy photons, we can also not rule this option out on
the basis of the current evidence.

\subsection{The SW Sex phenomenon explained (?)}

The structure of SW Sex as sketched above is capable of explaining many of the
classical SW Sex phenomena as we will outline here:
\begin{itemize}
\item Single peaked emission lines. We have shown that the emission
lines are formed at the hot-spot location. Since
they are formed in a single region the lines will also be single
peaked. 
\item Large phase shifts. The large phase shifts are naturally
explained by the origin of the lines in the hot-spot region. The precise
location of this region with respect to the center-of-mass will
determine the phase-shift observed in the emission lines, and can in
principle change with epoch.
\item Phase 0.5 absorption. Although this absorption is not detected
in our current observations it is relatively easy to imagine that
the absorption is due
to back-lighting of material overflowing the hot-spot region by the
hot-spot continuum radiation as seen at $\varphi\sim$0.5. 
If the mass-transfer rate from the
secondary is larger than observed here for SW Sex it is well possible
that a larger amount of the mass is transferred to the inner disk
which may cause the veiling. Back-lighting by the hot-spot continuum
radiation certainly explains the phase-dependence of the absorption
features which have been shown to have maximum depth around $\varphi
\sim$0.45, exactly when we see the hot-spot region from across the
disk (Szkody \&  Pich\'e, 1990). 
\item Shallow eclipses of the emission lines. Since the emission lines
are formed above the disk, they will show eclipses that are less deep
than that of the continuum.
\end{itemize}

Classically the SW Sex phenomenon was constrained to eclipsing systems
in the period range between 3 and 4 hours. Over the last few years
however, several systems at lower inclination or outside the 3-4 hr
orbital period range have been proposed as SW Sex stars, e.g. V795
Her (Casares et al., 1996), LS Pegasi (Mart\'{\i}nez-Pais et al., 1999;
Taylor et al., 1999), WX Arietis (Hellier et al., 1994), 
BT Mon (Smith et al., 1998) and V Per (this work). 
The scenario as given above does not depend on the
inclination angle, although we expect that at very low inclination
angles the transient absorption will become less strong. At
non-eclipsing inclination angles the phase 0.5 absorption will be 
caused by wind
material that is back-lit by either the hot spot, or the outer disk
region behind the hot spot, closer to the secondary. 
If the scenario outlined above holds true we expect that the
velocities at which the absorption components are found will become
larger with lower inclination, simply because we are looking more and
more directly into the wind. 

The attractiveness of this scenario over the others mentioned in Sect.\
\ref{sec:intro} is the fact that it is not only able to explain all the
observed feature of SW Sex stars, but especially the V-shaped Balmer
emission line light curves, which cannot be explained by any of the
other models. 

\begin{acknowledgements}
PJG is supported by the NWO Spinoza grant 08-0 to E.P.J. van
den Heuvel and a CfA fellowship. 
PJG wishes to thank the hospitality of Claudio Moreno and
the staff of the ING observatory during a number of visits. The INT is
operated on the Observatorio del Roque de los Muchachos on the island
of La Palma on behalf of the British PPARC and the Dutch NWO. 
\end{acknowledgements}

\end{document}